\documentclass[
  pdflatex,
  sn-mathphys-num,
  iicol,
  oneside
]{sn-jnl}

\usepackage{geometry}
\usepackage{etoolbox}
\usepackage{graphicx}%
\usepackage{multirow}%
\usepackage{amsmath,amssymb,amsfonts}%
\usepackage{amsthm}%
\usepackage{mathrsfs}%
\usepackage{xcolor}%
\usepackage{textcomp}%
\usepackage{manyfoot}%
\usepackage{booktabs}%
\usepackage{algorithm}%
\usepackage{algorithmicx}%
\usepackage{algpseudocode}%
\usepackage{listings}%
\usepackage{mathtools}
\usepackage{bm}
\usepackage{braket}
\usepackage{enumitem}
\usepackage[nameinlink,capitalise]{cleveref}
\usepackage{titlesec}

\titleformat{\section}[hang]
  {\normalfont\normalsize\bfseries}
  {\thesection}
  {0.6em}
  {}

\titleformat{\subsection}[hang]
  {\normalfont\small\bfseries}
  {\thesubsection}
  {0.55em}
  {}

\titleformat{\subsubsection}[hang]
  {\normalfont\small\bfseries}
  {\thesubsubsection}
  {0.5em}
  {}

\titlespacing*{\section}
  {0pt}
  {1.5ex plus 0.3ex minus 0.2ex}
  {0.6ex}

\titlespacing*{\subsection}
  {0pt}
  {0.7ex plus 0.2ex minus 0.1ex}
  {0.25ex}

\titlespacing*{\subsubsection}
  {0pt}
  {0.6ex plus 0.2ex minus 0.1ex}
  {0.2ex}

\makeatletter
\renewcommand{\Titlefont}{%
  \reset@font
  \fontsize{17bp}{22.5bp}%
  \bfseries
  \selectfont
  \titraggedcenter
}

\renewcommand{\Authorfont}{%
  \reset@font
  \fontsize{11bp}{13bp}%
  \bfseries
  \boldmath
  \selectfont
  \titraggedcenter
}

\renewcommand{\addressfont}{%
  \reset@font
  \fontsize{9bp}{11bp}%
  \selectfont
  \titraggedcenter
}

\patchcmd{\@maketitle}
  {Corresponding author(s). E-mail(s):}
  {Corresponding author. E-mail:}
  {}{}

\renewcommand{\@@address}[2][]{%
  \g@addto@macro\auaddress{%
    \stepcounter{affn}%
    \xdef\@currentlabel{\theaffn}%
    \jmkLabel{\theaffn}%
    {\textsuperscript{#1}#2\par}%
  }%
}
\makeatother

\patchcmd{\email}
  {;\ }
  {.\ }
  {}{\PackageWarning{sn-article}{Could not patch \string\email punctuation}}
\newcommand{\Eout}[1]{\mathcal{E}^{\mathrm{out}}_{#1}}
\newcommand{\Ein}[1]{\mathcal{E}^{\mathrm{in}}_{#1}}
\newcommand{\Damp}[1]{\mathcal{D}_{#1}}
\newcommand{\id}{\mathcal{I}}

\newcommand{\Ccoll}{C_{\mathrm{coll}}}

\newcommand{\qjsd}{J}
\newcommand{\proj}[1]{\ket{#1}\!\bra{#1}}
\newcommand{\degrad}{\succeq_{\mathrm{deg}}}

\theoremstyle{thmstyleone}
\newtheorem{theorem}{Theorem}
\newtheorem{corollary}[theorem]{Corollary}
\theoremstyle{thmstyletwo}

\theoremstyle{thmstylethree}

\begin{document}

\title[Opposite post-processing orders of fermionic horizon channels]
{Opposite post-processing orders of fermionic horizon channels
and their quantum-resource monotonicity}

\author[1]{\fnm{Shuhao} \sur{Li}}

\author*[1]{\fnm{Jin-Ming} \sur{Liu}}
\email{jmliu@phy.ecnu.edu.cn}

\affil[1]{%
  \orgdiv{School of Physics},
  \orgname{East China Normal University},
  \orgaddress{\city{Shanghai} \postcode{200241}, \country{China}}%
}

\abstract{Relativistic quantum-information studies in noninertial and black-hole settings often determine resource
behavior through explicit calculations for particular input states and
state functionals, leaving unclear whether the recurring monotonic trends
originate from those choices or from a common underlying structure. In this work, we formulate the effective single-mode fermionic
horizon transformation as a pair of complementary exterior and
interior quantum channels, corresponding respectively to the
physically accessible and inaccessible sectors, and establish exact
post-processing orders in opposite directions. As the relativistic channel parameter $q$ increases, the
exterior channel becomes progressively degraded, whereas the interior
channel is ordered in the reverse direction. 
These relations extend to arbitrary multipartite settings. 
Consequently, every state functional that is non-increasing under the corresponding intermediate maps is non-increasing in homogeneous exterior sectors and non-decreasing in homogeneous interior sectors. 
The framework therefore applies to broad classes of quantum resources and correlations, including entanglement and occupation-basis coherence monotones, optimized Bell-functional quantities, and contractive-divergence correlation measures.
We further numerically evaluate collective coherence based on the quantum Jensen–Shannon divergence (QJSD) in the Garfinkle--Horowitz--Strominger (GHS) dilaton-black-hole spacetime, illustrating the predicted homogeneous monotonicity.
The recurring trends are therefore traced to a common
channel-ordering structure, while the physical setting determines the
parameterization of \(q\) and the resource-theoretic monotonicity determines
which output-state quantities inherit the order.}


\maketitle
\small

\section{Introduction}

The integration of relativity and quantum physics has long been at the forefront of physics. Investigating how relativistic effects arising from different reference frames and spacetime geometries influence quantum information represents one branch of this broader frontier~\cite{peresQuantumInformationRelativity2004}.
Early studies of quantum field theory in curved spacetime and noninertial
reference frames revealed that the definitions of particles and vacuum are
not entirely independent of the observer~\cite{fullingNonuniquenessCanonicalField1973}.
Hawking demonstrated that black-hole horizons give rise to thermal
radiation~\cite{hawkingParticleCreationBlack1975}, 
while the Fulling--Davies--Unruh effect showed that uniformly
accelerated observers perceive the Minkowski vacuum as a thermal
environment~\cite{fullingNonuniquenessCanonicalField1973,daviesScalarProductionSchwarzschild1975,unruhNotesBlackholeEvaporation1976a}.
With the development of quantum-information theory, prior studies further
showed that spin--momentum coupling induced by Lorentz transformations
renders reduced spin entropy and entanglement reference-frame
dependent~\cite{peresQuantumEntropySpecial2002,gingrichQuantumEntanglementMoving2002}.
Subsequent studies investigated quantum teleportation involving a
uniformly accelerated observer~\cite{alsingTeleportationUniformlyAccelerated2003},
demonstrated the observer dependence of field-mode entanglement in
noninertial reference frames~\cite{fuentes-schullerAliceFallsBlack2005a},
and further extended the analysis to Dirac fields~\cite{
alsingEntanglementDiracFields2006}.
These studies connected the observer dependence of quantum field theory
with entanglement, communication tasks, and resource descriptions in
quantum information, thereby laying the early foundations of relativistic
quantum information and promoting its gradual extension from information
transformations under relativity to quantum resources and information
tasks in noninertial frames, horizon settings, and curved spacetimes~\cite{alsingObserverdependentEntanglement2012,huRelativisticQuantumInformation2012,martin-martinezEntanglementCurvedSpacetimes2014}.
Research in this field broadly follows two directions:
one examines the effects of relativistic settings on quantum information
and quantum resources, while the other employs quantum information theory and
detector-based methods to investigate quantum fields and spacetime.
The problem considered in this work belongs to the former category.

In recent years, a substantial body of research has explored the effects
of relativistic settings on quantum information and quantum resources
across a broad range of physical scenarios~\cite{adessoContinuousvariableEntanglementSharing2007,panHawkingRadiationEntanglement2008,zhangEntropicUncertaintyRelation2018,lianQuantumFisherInformation2021,yaoHierarchicalRelationshipNonlocal2025,jiNonlocalCorrelationQuantum2025,liQuantumnessEntropicUncertainty2022,zhangHawkingEffectCan2023,liuQuantumCoherenceQuantum2024,liBosonicFermionicCoherence2024,huangGenuineEntanglementMultievent2025,nasresfahaniReductionEntanglementDegradation2011,geQuantumEntanglementTeleportation2008a,wangEntanglementRedistributionSchwarzschild2010a,wangQuantumDiscordMeasurementinduced2014a,liuQuantumPropertiesFermionic2023,liEntropicUncertaintyCoherence2026}.
A widely used approach is to select a multipartite initial state, employ
relativistic field-mode transformations to describe correlations between
modes associated with different regions or observers, retain the target
modes while tracing out the complementary modes, and then examine how the
quantum-information properties and quantum resources of the resulting
reduced state vary with relativistic or geometric parameters including, among others,
acceleration, Hawking temperature, the dilaton parameter, and
spacetime dimensionality~\cite{wangClassicalCorrelationQuantum2010,wangEntanglementRedistributionSchwarzschild2010a,wangQuantumDiscordMeasurementinduced2014a,geQuantumEntanglementTeleportation2008a,liEntropicUncertaintyCoherence2026,liuQuantumPropertiesFermionic2023}.
Across this diverse literature, studies of homogeneous exterior (physically accessible) or interior (physically inaccessible) sectors have repeatedly reported one or both aspects of a
recurring pattern: as the effective degree of horizon excitation
increases, quantum resources in physically accessible sectors are often
found to decrease monotonically, whereas resources in the corresponding
physically inaccessible sectors may be generated or increase
monotonically~\cite{alsingEntanglementDiracFields2006,liQuantumnessEntropicUncertainty2022,zhangHawkingEffectCan2023,liuQuantumCoherenceQuantum2024,liBosonicFermionicCoherence2024,huangGenuineEntanglementMultievent2025,nasresfahaniReductionEntanglementDegradation2011,geQuantumEntanglementTeleportation2008a,wangEntanglementRedistributionSchwarzschild2010a,wangQuantumDiscordMeasurementinduced2014a,liuQuantumPropertiesFermionic2023,wangClassicalCorrelationQuantum2010,wangQuantumEntanglementDirac2010,xuProbingQuantumCorrelation2014,hePropertyVariousCorrelation2015,liuReversingQuantumResource2025,wuMaximallyEntangledStates2025,liEnhancedRobustnessNonmaximal2026,liMultiqubitCoherenceMixed2026a,tengBosonicFermionicMutual2026,xuHowHawkingEffect2014,asghariEntanglementDegradationPresence2018,haddadiQuantumnessSchwarzschildBlack2024,miGenuineFourpartiteBell2025}.

Although similar trends have repeatedly appeared in different studies, it remains insufficiently and explicitly understood whether they share a unified structural origin. The studies reporting this recurring pattern typically obtain these results through calculations tailored to specific initial states, spacetime backgrounds, and resource measures, and broadly interpret them as follows: the Hawking effect degrades quantum resources in physically accessible regions while generating or enhancing the corresponding resources in physically inaccessible regions, thereby inducing a redistribution of quantum resources across the horizon. 
However, this qualitative description alone is insufficient to explain why
similar evolutionary trends repeatedly arise for different initial states,
multipartite configurations, and resource measures. This naturally gives
rise to a more fundamental question: do these seemingly model- and
measure-dependent phenomena originate from a common underlying mechanism?
To the best of our knowledge, this question, which the present work is precisely devoted to addressing, has not yet been systematically investigated.

In this paper, we first prove that the homogeneous exterior and homogeneous interior channels induced by the fermionic mode transformation and by tracing over the respective complementary outputs possess oppositely directed post-processing orders for two componentwise ordered sets of relativistic channel parameters. This structure is independent of the specific input state and can be extended to arbitrary multipartite systems and arbitrary subsystems affected by the channels. By further combining the monotonicity of state functionals under the intermediate maps, the preservation of positive operator-valued measure (POVM) validity under measurement pullback, and the covariance between the channels and the marginal structures, we find that broad classes of entanglement and
occupation-basis coherence monotones, optimized Bell-functional
quantities, and contractive-divergence correlation measures are
monotonically non-increasing in homogeneous physically accessible
sectors and monotonically non-decreasing in homogeneous physically
inaccessible sectors. These results show that, within the homogeneous sectors, the repeatedly observed characteristic trends originate from the opposite post-processing orders of the two classes of channels and the simple properties satisfied by the corresponding state functionals. Finally, using collective coherence based on the quantum Jensen–Shannon divergence (QJSD)~\cite{majteyJensenShannonDivergenceMeasure2005,radhakrishnanDistributionQuantumCoherence2016,radhakrishnanBasisindependentQuantumCoherence2019}
in the Garfinkle–Horowitz–Strominger (GHS) spacetime~\cite{garfinkleChargedBlackHoles1991a}
as an example, we explicitly demonstrate the application and scope of our results.

The remainder of this paper is organized as follows.
In Sec.~\ref{sec:fermionic-channels}, we establish the opposite
post-processing orders of the exterior and interior horizon channels.
In Sec.~\ref{sec:resources}, we derive their consequences
for resource measures and related state functionals.
In Sec.~\ref{sec:ghs}, we apply the framework to collective
coherence in the GHS spacetime.
Finally, Sec.~\ref{sec:discussion} discusses the scope and
limitations of the approach and summarizes the main conclusions.

\section{Fermionic horizon channels and their opposite post-processing orders}
\label{sec:fermionic-channels}
In this section we start with a fermionic mode transformation commonly used in a class of relativistic settings. By tracing over the two complementary outputs separately, we construct the single-mode exterior and interior channels, derive their matrix actions on a general input state and the corresponding Kraus representations, and verify that both channels are completely positive trace-preserving (CPTP) maps. We then explicitly construct intermediate maps to prove that, for two ordered values of the relativistic channel parameter, the exterior and interior channels obey post-processing orders in opposite directions. On this basis, we extend the result to multipartite homogeneous-exterior and homogeneous-interior sectors, and discuss the scope of the present post-processing construction in mixed exterior–interior sectors.

\subsection{Fermionic mode transformation and relativistic parameterization}
Within the widely adopted effective single-mode, two-level occupation-number
description for Dirac fields in noninertial frames and
near spacetime horizons, the transformation for a chosen fermionic mode reads
\cite{alsingEntanglementDiracFields2006,wangQuantumEntanglementDirac2010,wangEntanglementRedistributionSchwarzschild2010a,wangQuantumDiscordMeasurementinduced2014a}
\begin{equation}
\begin{aligned}
  \ket{0}
  &\longmapsto
  \sqrt{1-q}\,
  \ket{0}_{\mathrm{out}}\ket{0}_{\mathrm{in}}
  +
  \sqrt{q}\,
  \ket{1}_{\mathrm{out}}\ket{1}_{\mathrm{in}},
  \\
  \ket{1}
  &\longmapsto
  \ket{1}_{\mathrm{out}}\ket{0}_{\mathrm{in}}.
\end{aligned}
\label{eq:fermionic-mode-transformation}
\end{equation}
The labels ``out'' and ``in'' denote the two causally separated mode
sectors appropriate to the physical setting. The vacuum is represented
as a superposition of the jointly unoccupied and jointly occupied
configurations of these sectors, whereas the one-particle state has the
exterior mode occupied and the interior mode unoccupied. Accordingly,
$q$ is the occupation weight of the jointly occupied component generated
by the horizon transformation. Within this effective two-level
description, Eq.~\eqref{eq:fermionic-mode-transformation} defines an
isometric embedding for $0\leq q\leq1$. Because the two output sectors are causally separated, an observer
restricted to either sector has no access to the complementary output.
Tracing over that inaccessible sector therefore turns the effective
two-output transformation into the exterior or interior quantum
channel constructed in the next subsection.

The dependence of $q$ on the underlying physical parameters is
scenario dependent. For a uniformly accelerated observer with proper
acceleration $a$,
$q=[1+\exp(2\pi\omega/a)]^{-1}$
\cite{alsingEntanglementDiracFields2006}. For a fermionic mode of
frequency $\omega$ near a Schwarzschild black hole of mass $M$,
$q=[1+\exp(8\pi M\omega)]^{-1}$
\cite{wangQuantumEntanglementDirac2010,wangEntanglementRedistributionSchwarzschild2010a}.
For the GHS dilaton black hole considered below,
$q=[1+\exp(8\pi(M-D)\omega)]^{-1}$, where $D$ is the
dilaton parameter
\cite{wangQuantumDiscordMeasurementinduced2014a}.
The expressions above are written in natural units,
\(\hbar=c=k_{\mathrm B}=G=1\). In all three cases, \(q\) has the common
form \(q=[1+\exp(x)]^{-1}\), where \(x>0\) for the positive-frequency,
finite-parameter, and nonextremal settings considered here. It therefore
follows that $0<q<\frac{1}{2}$,
with the endpoints approached only in limiting regimes.
Related effective two-level transformations have also been employed
for higher-dimensional and rotating black holes,
Einstein--Gauss--Bonnet gravity, and cosmological horizons
\cite{geQuantumEntanglementTeleportation2008a,nasresfahaniReductionEntanglementDegradation2011,bhattacharyaDiracFermionCosmological2020a}.
Thus, whenever the relevant mode transformation reduces to
Eq.~\eqref{eq:fermionic-mode-transformation}, the spacetime geometry
determines the parameterization of $q$, while the channel-ordering
argument developed below depends only on the common algebraic form.

Throughout this work, we adopt the effective mode-qubit description
widely used in relativistic quantum information. Within this
description, each selected fermionic occupation mode is treated as an
effective two-dimensional subsystem, and multimode states and channels
are represented using ordinary tensor products and partial traces. The
limitations of the single-mode reduction and the subtleties associated
with strict fermionic subsystem structures are discussed in
Sec.~\ref{sec:discussion}.

\subsection{Induced exterior and interior channels and their opposite post-processing orders}
\label{sec:smpopo}

We now construct explicitly the two reduced channels described in the
preceding subsection. The reduced dynamics obtained by
tracing over the inaccessible ``in'' output has previously been
formulated as a fermionic Unruh noise channel
\cite{omkarUnruhEffectInterpreted2016}. Here, we place the two
complementary reductions of
Eq.~\eqref{eq:fermionic-mode-transformation} on an equal footing.

A general input state in the occupation-number basis can be written as
\begin{equation}
  \rho
  =
  a\ket{0}\!\bra{0}
  +
  b\ket{0}\!\bra{1}
  +
  b^{*}\ket{1}\!\bra{0}
  +
  d\ket{1}\!\bra{1},
\end{equation}
where
$a=\langle 0|\rho|0\rangle$,
$d=\langle 1|\rho|1\rangle$, and
$b=\langle 0|\rho|1\rangle$, with
$a+d=1$.
Since both the mode transformation and the partial trace are linear,
it is sufficient to evaluate the four operator-basis elements
separately.
For compactness, we write
$\ket{ij}\equiv\ket{i}_{\mathrm{out}}\ket{j}_{\mathrm{in}}$,
with the ``out'' occupation label appearing first and the ``in''
occupation label second. For example,
Eq.~\eqref{eq:fermionic-mode-transformation} maps the coherence
operator as
\begin{equation}
  \ket{0}\!\bra{1}
  \longmapsto
  \sqrt{1-q}\,\ket{00}\!\bra{10}
  +
  \sqrt{q}\,\ket{11}\!\bra{10}.
\end{equation}
Under the partial trace, only terms with matching occupation labels in
the traced-out output survive. Thus, tracing over the ``out'' output
in the expression above retains only the second term and gives
$\sqrt{q}\,\ket{1}_{\mathrm{in}}\!\bra{0}$.
The remaining basis elements and the complementary reduction are obtained analogously.

By linearity, tracing over the ``in'' output yields the exterior
channel
\begin{equation}
  \Eout{q}(\rho)
  =
  \begin{pmatrix}
    (1-q)a & \sqrt{1-q}\,b\\
    \sqrt{1-q}\,b^{*} & d+qa
  \end{pmatrix},
  \label{eq:exterior-channel}
\end{equation}
whereas tracing over the ``out'' output yields the complementary
interior channel
\begin{equation}
  \Ein{q}(\rho)
  =
  \begin{pmatrix}
    d+(1-q)a & \sqrt{q}\,b^{*}\\
    \sqrt{q}\,b & qa
  \end{pmatrix}.
  \label{eq:interior-channel}
\end{equation}
In Eqs.~\eqref{eq:exterior-channel} and
\eqref{eq:interior-channel}, the matrices are expressed in the
occupation-number basis of the retained output.

A Kraus representation of the exterior channel is given by
\begin{equation}
  K_{0}^{\mathrm{out}}
  =
  \begin{pmatrix}
    \sqrt{1-q}&0\\
    0&1
  \end{pmatrix},
  \qquad
  K_{1}^{\mathrm{out}}
  =
  \begin{pmatrix}
    0&0\\
    \sqrt{q}&0
  \end{pmatrix}.
  \label{eq:exterior-kraus}
\end{equation}
Similarly, a Kraus representation of the interior channel is given by
\begin{equation}
  K_{0}^{\mathrm{in}}
  =
  \begin{pmatrix}
    \sqrt{1-q}&0\\
    0&0
  \end{pmatrix},
  \qquad
  K_{1}^{\mathrm{in}}
  =
  \begin{pmatrix}
    0&1\\
    \sqrt{q}&0
  \end{pmatrix}.
  \label{eq:interior-kraus}
\end{equation}
Both Kraus representations satisfy the completeness condition and
therefore define CPTP maps~\cite{nielsenQuantumComputationQuantum2010}.
Hence, $\Eout{q}$ and $\Ein{q}$ form a complementary pair of channels
induced by the same effective two-output mode transformation. As $q$
varies over its physical range, these maps form the exterior and
interior channel families, respectively.

\paragraph*{Post-processing preorder.}
To compare the members of these channel families at different values of $q$, we use the post-processing preorder for quantum channels~\cite{buscemiDegradableChannelsLess2016,jencovaGeneralTheoryComparison2021}.
Let $\mathcal M$ and $\mathcal N$ be quantum channels with the same
input space. We write
\begin{equation*}
  \mathcal M\degrad\mathcal N
  \Longleftrightarrow
  \exists\,\Theta\ \text{CPTP}:\ 
  \mathcal N=\Theta\circ\mathcal M.
\end{equation*}
In this case, $\mathcal N$ is a degraded version of $\mathcal M$, or
equivalently, it can be obtained from $\mathcal M$ by the
post-processing channel $\Theta$. Operationally, the output of
$\mathcal N$ can be simulated from the output of $\mathcal M$ by
applying the additional channel $\Theta$. The direction of this
preorder need not coincide with the numerical ordering of the
parameters labeling the channel family.

We first consider the exterior channel family. For
$0<q_1<q_2<1/2$, set $p=\frac{q_2-q_1}{1-q_1}$.
Then $0<p<1$. Using $1-q_2=(1-p)(1-q_1)$ and $q_2=q_1+p(1-q_1)$,
direct substitution into Eq.~\eqref{eq:exterior-channel} verifies that,
for every input state $\rho$, we have 
$
\Eout{p}\!\left[\Eout{q_1}(\rho)\right]
=
\Eout{q_2}(\rho).
$
Therefore,
$
  \Eout{q_2}
  =
  \Eout{p}\circ\Eout{q_1},
$
and hence
\begin{equation}
  q_2>q_1
  \quad\Longrightarrow\quad
  \Eout{q_1}\degrad\Eout{q_2}.
  \label{eq:exterior-order}
\end{equation}
Thus, the exterior-channel family decreases in the degradation preorder as $q$ increases.

We next consider the interior channel family. Consider the
amplitude-damping channel with parameter $\eta\in[0,1]$
\cite{nielsenQuantumComputationQuantum2010},
\begin{equation}
  \Damp{\eta}
  \begin{pmatrix}
    a&b\\
    b^{*}&d
  \end{pmatrix}
  =
  \begin{pmatrix}
    a+(1-\eta)d & \sqrt{\eta}\,b\\
    \sqrt{\eta}\,b^{*} & \eta d
  \end{pmatrix}.
  \label{eq:amplitude-damping}
\end{equation}
Its Kraus operators are
\begin{equation}
  L_0=
  \begin{pmatrix}
    1&0\\
    0&\sqrt{\eta}
  \end{pmatrix},
  \qquad
  L_1=
  \begin{pmatrix}
    0&\sqrt{1-\eta}\\
    0&0
  \end{pmatrix}.
  \label{eq:amplitude-damping-kraus}
\end{equation}

For $0<q_1<q_2<1/2$, set $\eta=\frac{q_1}{q_2}$.
Then $0<\eta<1$. Direct substitution of
Eq.~\eqref{eq:interior-channel} into
Eq.~\eqref{eq:amplitude-damping} verifies that, for every input state
$\rho$, we have
$
\Damp{\eta}\!\left[\Ein{q_2}(\rho)\right]
=
\Ein{q_1}(\rho).
$
Therefore,
$
  \Ein{q_1}
  =
  \Damp{\eta}\circ\Ein{q_2},
$
and hence
\begin{equation}
  q_2>q_1
  \quad\Longrightarrow\quad
  \Ein{q_2}\degrad\Ein{q_1}.
  \label{eq:interior-order}
\end{equation}
Thus, the interior-channel family increases in the degradation preorder
as $q$ increases.

Equations~\eqref{eq:exterior-order} and
\eqref{eq:interior-order} constitute the central single-mode
structural result of this work: the two complementary channel families
induced by the same effective fermionic mode transformation obey
opposite post-processing orders as the parameter $q$ increases.
Although these channel relations are simple, when combined with
appropriate properties of state functionals, they reveal a common
structure underlying a broad class of previously state-, measure-, and
spacetime-specific calculations. Before turning to these consequences,
we first extend the channel ordering to multipartite systems.

\subsection{Multipartite extension and the scope of the ordering}
\label{sec:mixed-sectors}

Consider an arbitrary $N$-partite input state
$\rho_{A_1A_2\cdots A_N}$. Suppose that $n\leq N$ selected fermionic
mode subsystems are affected by the effective mode transformation in
Eq.~\eqref{eq:fermionic-mode-transformation}. We denote the set of
affected modes by $\mathcal H$ and the remaining unaffected subsystems
by $\mathcal H^c$. In a black-hole setting, for example, the affected
subsystems may represent field modes associated with observers near
the horizon, for which an out--in mode decomposition is required,
while the other subsystems remain unaffected.

For each $i\in\mathcal H$, the local mode transformation produces an
``out'' output and an ``in'' output. We assume that these
transformations act locally and independently on the selected modes and
introduce no additional coupling between different mode subsystems.
This local product construction is widely used in multipartite
relativistic quantum-information calculations involving accelerated
observers and fermionic modes near spacetime horizons.
The corresponding multipartite transformation is therefore obtained
by tensoring the local transformations, together with the identity
operation on the unaffected subsystems.

Retaining all ``out'' outputs and tracing over the complementary
``in'' outputs gives the homogeneous exterior channel
\begin{subequations}
\begin{equation}
  \Phi_{\bm q}^{\mathrm{out}}
  =
  \left(
    \bigotimes_{i\in\mathcal H}
    \Eout{q_i}
  \right)
  \otimes
  \id_{\mathcal H^c},
\end{equation}
whereas retaining all ``in'' outputs and tracing over the complementary
``out'' outputs gives the homogeneous interior channel
\begin{equation}
  \Phi_{\bm q}^{\mathrm{in}}
  =
  \left(
    \bigotimes_{i\in\mathcal H}
    \Ein{q_i}
  \right)
  \otimes
  \id_{\mathcal H^c}.
\end{equation}
\end{subequations}
Here, ``homogeneous'' refers only to the common choice of retained
output sector: all affected modes are represented either by their
exterior outputs or by their interior outputs. The local parameters
$q_i$ need not be equal, and we collect them into the vector
$\bm q=(q_i)_{i\in\mathcal H}$. The single-mode post-processing
relations established in Sec.~\ref{sec:smpopo} can then be applied
independently to each affected mode, leading to the following
componentwise multipartite ordering.

\begin{theorem}[Componentwise multipartite ordering]
\label{thm:componentwise-multipartite-ordering}

Suppose that \(0<q_i\leq q_i'<1/2\) for every
\(i\in\mathcal H\). For each \(i\in\mathcal H\), define
\(p_i=(q_i'-q_i)/(1-q_i)\) and \(\eta_i=q_i/q_i'\).
Then the corresponding multipartite exterior and interior channels
satisfy the following post-processing relations:
\begin{subequations}
\label{eq:multipartite-postprocessing}

\begin{equation}
\resizebox{0.8\linewidth}{!}{$\displaystyle
  \Phi_{\bm q'}^{\mathrm{out}}
  =
  \mathcal V_{\bm q'\leftarrow\bm q}^{\mathrm{out}}
  \circ \Phi_{\bm q}^{\mathrm{out}},
  \quad
  \mathcal V_{\bm q'\leftarrow\bm q}^{\mathrm{out}}
  =
  \left(
    \bigotimes_{i\in\mathcal H}\Eout{p_i}
  \right)
  \otimes \id_{\mathcal H^c}
$}
\label{eq:multipartite-postprocessing-out}
\end{equation}

\begin{equation}
\resizebox{0.8\linewidth}{!}{$\displaystyle
  \Phi_{\bm q}^{\mathrm{in}}
  =
  \mathcal V_{\bm q\leftarrow\bm q'}^{\mathrm{in}}
  \circ \Phi_{\bm q'}^{\mathrm{in}},
  \;
  \mathcal V_{\bm q\leftarrow\bm q'}^{\mathrm{in}}
  =
  \left(
    \bigotimes_{i\in\mathcal H}\Damp{\eta_i}
  \right)
  \otimes \id_{\mathcal H^c}
$}
\label{eq:multipartite-postprocessing-in}
\end{equation}

\end{subequations}
Equivalently, the two homogeneous channel families satisfy the opposite
post-processing preorders
\begin{equation}
  \Phi_{\bm q}^{\mathrm{out}}
  \degrad
  \Phi_{\bm q'}^{\mathrm{out}},
  \qquad
  \Phi_{\bm q'}^{\mathrm{in}}
  \degrad
  \Phi_{\bm q}^{\mathrm{in}}.
\end{equation}
\end{theorem}

For every component with $q_i=q_i'$, the corresponding intermediate map is the identity. For the remaining components, the result follows by tensoring the single-mode composition relations in Eqs.~\eqref{eq:exterior-order} and \eqref{eq:interior-order} over all affected modes.

For a mixed exterior--interior configuration, partition the affected
set as
$\mathcal H=\mathcal H_{\mathrm{out}}\mathbin{\dot\cup}
\mathcal H_{\mathrm{in}}$,
where both subsets are nonempty. The corresponding mixed channel is
\begin{equation}
  \Phi_{\bm q}^{\mathrm{mix}}
  =
  \left(
    \bigotimes_{i\in\mathcal H_{\mathrm{out}}}
    \Eout{q_i}
  \right)
  \otimes
  \left(
    \bigotimes_{j\in\mathcal H_{\mathrm{in}}}
    \Ein{q_j}
  \right)
  \otimes
  \id_{\mathcal H^c}.
\end{equation}
For two componentwise ordered parameter vectors
$\bm q\leq\bm q'$, the exterior factors are post-processed from
$\bm q$ to $\bm q'$, whereas the interior factors require the opposite
post-processing direction. Consequently, the local intermediate maps used in Theorem~\ref{thm:componentwise-multipartite-ordering} cannot
in general be combined into a single product post-processing channel
in either direction. The homogeneous ordering theorem therefore does not apply
directly to mixed exterior--interior configurations, and the
monotonicity of quantum resources in such sectors is not guaranteed by
the present ordering argument.

\section{From channel ordering to monotonicity of state functionals}
\label{sec:resources}

In the previous section, we established the opposite post-processing
orders of the exterior and interior channel families in multipartite
homogeneous sectors.
We now combine these channel-level orders with the monotonicity
of specific state functionals under the corresponding intermediate maps
to determine how the resulting resource measures and operational
quantities vary with the effective channel parameter \(q\).
For an arbitrary multipartite input state $\rho_0$, define
\begin{equation}
  \rho_{\mathrm{out}}(\bm q)
  =
  \Phi_{\bm q}^{\mathrm{out}}(\rho_0),
  \qquad
  \rho_{\mathrm{in}}(\bm q)
  =
  \Phi_{\bm q}^{\mathrm{in}}(\rho_0).
\end{equation}
With this notation,
Theorem~\ref{thm:componentwise-multipartite-ordering}
yields the following functional ordering.

\begin{corollary}[Ordering of state functionals for homogeneous horizon channels]
\label{cor:homogeneous-resource-ordering}
Let $\rho_0$ be an arbitrary multipartite input state,
let $\mathcal H$ be an arbitrary set of affected modes,
and let $\bm q$ and $\bm q'$ satisfy the componentwise ordering condition
of Theorem~\ref{thm:componentwise-multipartite-ordering}.
For the exterior sector, suppose that the state functional $R$
is non-increasing under the corresponding intermediate map, namely,
$
  R\!\left[
    \mathcal V_{\bm q' \leftarrow \bm q}^{\mathrm{out}}(\sigma)
  \right]
  \leq
  R(\sigma)
$
for every state $\sigma$ on the output space. Then
\begin{subequations}
\begin{equation}
  R[\rho_{\mathrm{out}}(\bm q')]
  \leq
  R[\rho_{\mathrm{out}}(\bm q)].
\end{equation}
For the interior sector, suppose instead that
$
  R\!\left[
    \mathcal V_{\bm q \leftarrow \bm q'}^{\mathrm{in}}(\sigma)
  \right]
  \leq
  R(\sigma)
$
for every state $\sigma$ on the output space. Then
\begin{equation}
  R[\rho_{\mathrm{in}}(\bm q')]
  \geq
  R[\rho_{\mathrm{in}}(\bm q)].
\end{equation}
\end{subequations}
\end{corollary}

\begin{figure*}[t]
  \centering
  \includegraphics[width=\linewidth]{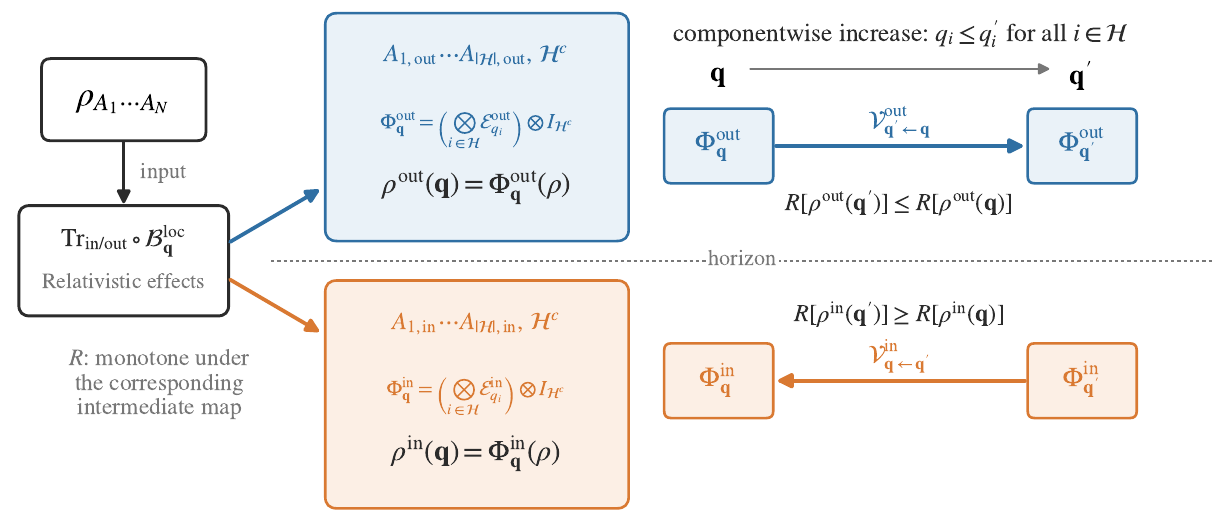}
  \caption{Schematic representation of the channel-to-functional ordering
  mechanism in multipartite homogeneous horizon sectors.}
  \label{fig:channel-functional-ordering}
\end{figure*}

The two inequalities follow directly from the post-processing relations
in Theorem~\ref{thm:componentwise-multipartite-ordering}
and the assumed non-increase of $R$
under the corresponding intermediate maps.
The channel-to-functional ordering mechanism is summarized schematically
in Fig.~\ref{fig:channel-functional-ordering}.

In what follows, the subsections~\ref{sec:entanglement-monotones}--\ref{sec:contractive-correlations} establish the required monotonicity
for several classes of state functionals
using class-dependent structural arguments.
Mixed exterior--interior sectors are excluded from this unified result
because they do not in general admit a common post-processing direction.

Corollary~\ref{cor:homogeneous-resource-ordering}
separates two logically independent ingredients:
the relativistic model determines the post-processing order
of the effective channel family,
whereas the resource-theoretic or operational framework
determines which state functionals are non-increasing
under the corresponding intermediate maps.
Together, these ingredients determine the variation
of such functionals with the effective channel parameter \(q\).
In a specific physical setting,
the relation between \(q\) and quantities such as acceleration,
black-hole parameters, or mode frequency
translates this ordering into trends
with respect to the corresponding physical parameters.
This framework reduces the question raised in the Introduction to
identifying which state functionals are monotone under the relevant
intermediate maps.

\subsection{Entanglement monotones}
\label{sec:entanglement-monotones}

In this subsection we begin with entanglement functionals. For either homogeneous sector,
let $\mathcal V$ denote the corresponding multipartite intermediate
map established in
Theorem~\ref{thm:componentwise-multipartite-ordering}. Consider an arbitrary fixed multipartite partition
$\mathcal P=B_1|B_2|\cdots|B_m$, where each block $B_j$ may contain
one or more elementary subsystems.
Since $\mathcal V$ acts independently on the elementary subsystems,
its local factors can be regrouped according to this partition as
$
  \mathcal V
  =
  \mathcal V_{B_1}
  \otimes
  \mathcal V_{B_2}
  \otimes\cdots\otimes
  \mathcal V_{B_m},
$
where $\mathcal V_{B_j}$ denotes the tensor product of the elementary
local maps acting on the subsystems contained in $B_j$,
including identity maps on unaffected subsystems.
Thus, relative to any fixed bipartite or multipartite coarse-grained
partition $\mathcal P$, the intermediate map is a deterministic local
CPTP map of product form.

Accordingly, let $E_{\mathcal P}$ be an entanglement functional
defined relative to the fixed partition $\mathcal P$ and
non-increasing under deterministic local CPTP maps of product form
\cite{vidalEntanglementMonotones2000,horodeckiQuantumEntanglement2009}.
Therefore, whenever $\bm q$ and $\bm q'$ satisfy the componentwise
ordering condition of
Theorem~\ref{thm:componentwise-multipartite-ordering},
Corollary~\ref{cor:homogeneous-resource-ordering} yields
\begin{subequations}
\begin{equation}
\begin{aligned}
E_{\mathcal P}\!\left[\rho_{\mathrm{out}}(\bm q')\right]
&\leq
E_{\mathcal P}\!\left[\rho_{\mathrm{out}}(\bm q)\right],
\end{aligned}
\end{equation}
\begin{equation}
\begin{aligned}
E_{\mathcal P}\!\left[\rho_{\mathrm{in}}(\bm q')\right]
&\geq
E_{\mathcal P}\!\left[\rho_{\mathrm{in}}(\bm q)\right].
\end{aligned}
\end{equation}
\end{subequations}

These relations apply, in particular, to any fixed bipartition
$X|\bar X$.
Representative examples include the negativity and logarithmic
negativity
\cite{vidalComputableMeasureEntanglement2002,plenioLogarithmicNegativityFull2005},
the relative entropy of entanglement
\cite{vedralQuantifyingEntanglement1997,vedralEntanglementMeasuresPurification1998},
the entanglement of formation
\cite{bennettMixedstateEntanglementQuantum1996},
squashed entanglement
\cite{christandlSquashedEntanglementAdditive2004},
and, for two-qubit systems, the concurrence
\cite{hillEntanglementPairQuantum1997,woottersEntanglementFormationArbitrary1998}.
The same reasoning extends to genuine multipartite entanglement,
because deterministic local CPTP maps of product form preserve the
set of biseparable states. Representative examples include the genuine
multipartite concurrence
\cite{maMeasureGenuineMultipartite2011},
the genuine multiparticle negativity based on the PPT-mixture
framework
\cite{jungnitschTamingMultiparticleEntanglement2011,hofmannAnalyticalCharacterizationGenuine2014},
and suitable geometric multipartite entanglement measures
\cite{senChannelCapacitiesEntanglement2010,dasGeneralizedGeometricMeasure2016}.
This list of entanglement measures is not exhaustive; the same
exterior and interior ordering holds for any entanglement functional
that is non-increasing under the corresponding intermediate local
maps.

A transparent illustration of these ordering relations is provided by
the Schwarzschild entanglement-redistribution analysis of
Ref.~\cite{wangEntanglementRedistributionSchwarzschild2010a}.
For a fermionic mode of fixed frequency $\omega$, the effective channel
parameter
$q=[1+\exp(\frac{\omega}{T_{\mathrm H}})]^{-1}$
increases monotonically with the Hawking temperature $T_{\mathrm H}$.
The exterior ordering therefore predicts that the entanglement across the accessible bipartition $A|I$ is non-increasing with $T_{\mathrm H}$,
whereas the interior ordering predicts that the entanglement across the inaccessible bipartition $A|II$ is non-decreasing.
These predictions agree with the concurrence and entanglement-of-formation
results obtained in the cited Schwarzschild study~\cite{wangEntanglementRedistributionSchwarzschild2010a}.
The same channel-ordering mechanism provides a common structural
explanation for representative homogeneous-sector entanglement trends
reported for Dirac fields in noninertial frames and in Schwarzschild
and GHS black-hole spacetimes, in both bipartite and genuine
multipartite settings
\cite{alsingEntanglementDiracFields2006,xuHowHawkingEffect2014,qiangGenuineMultipartiteConcurrence2018,liuReversingQuantumResource2025}.
Related homogeneous exterior trends have also been reported in
multi-horizon backgrounds
\cite{huangGenuineEntanglementMultievent2025}
and for non-maximally entangled bipartite and multipartite states
\cite{Yang2026Advantages,liEnhancedRobustnessNonmaximal2026}.
These references are representative rather than exhaustive. More
generally, the present framework covers homogeneous exterior- and
interior-sector entanglement trends whenever the parameters are
componentwise ordered and the chosen entanglement functional is
monotone under the corresponding intermediate local CPTP maps.
It determines the monotonic direction, but not the numerical value, any possible sudden-death threshold, or the mixed-sector behavior.

\subsection{Occupation-basis coherence monotones}

We next consider occupation-basis coherence within the resource theory
of quantum coherence
\cite{baumgratzQuantifyingCoherence2014,streltsovColloquiumQuantumCoherence2017}.
We fix the occupation-number basis $\{|0\rangle,|1\rangle\}$ for each
mode and the corresponding product basis for the multipartite system.
A channel is a strictly incoherent operation (SIO) if it admits a
Kraus representation in which every Kraus operator and its adjoint are
incoherent; equivalently, each Kraus operator has at most one nonzero
entry in every row and every column
\cite{winterOperationalResourceTheory2016,yadinQuantumProcessesWhich2016}.
For the single-mode exterior and interior post-processing relations,
the intermediate maps are respectively $\Eout{p}$ and $\Damp{\eta}$,
with Kraus operators given in
Eqs.~\eqref{eq:exterior-kraus} and
\eqref{eq:amplitude-damping-kraus}.
These Kraus operators satisfy the SIO criterion.
Since tensor products with identity channels preserve this property,
the multipartite intermediate maps implementing the homogeneous
exterior and interior post-processing relations are also SIO relative
to the product occupation-number basis.

Let $C_{\mathrm{occ}}$ be an occupation-basis coherence functional
that is non-increasing under SIOs.
Since the multipartite intermediate maps established above are SIO,
Corollary~\ref{cor:homogeneous-resource-ordering} implies, for
componentwise ordered parameter vectors $\bm q\leq\bm q'$,
\begin{subequations}
\begin{equation}
\begin{aligned}
C_{\mathrm{occ}}
\!\left[\rho_{\mathrm{out}}(\bm q')\right]
\leq
C_{\mathrm{occ}}
\!\left[\rho_{\mathrm{out}}(\bm q)\right],
\end{aligned}
\end{equation}
\begin{equation}
\begin{aligned}
C_{\mathrm{occ}}
\!\left[\rho_{\mathrm{in}}(\bm q')\right]
\geq
C_{\mathrm{occ}}
\!\left[\rho_{\mathrm{in}}(\bm q)\right].
\end{aligned}
\end{equation}
\end{subequations}
Thus, occupation-basis coherence is non-increasing in homogeneous
exterior sectors and non-decreasing in homogeneous interior sectors.

Standard examples include the $l_1$-norm coherence and the
relative entropy of coherence
\cite{baumgratzQuantifyingCoherence2014},
the robustness of coherence
\cite{napoliRobustnessCoherenceOperational2016},
and the coherence of formation
\cite{winterOperationalResourceTheory2016}.
These examples are representative rather than exhaustive; the same
ordering applies to every occupation-basis coherence functional that
is non-increasing under SIO.

A representative illustration is provided by the $l_1$-norm coherence
of multipartite GHZ and W states in the GHS dilaton spacetime
\cite{liBosonicFermionicCoherence2024}.
For the fermionic field, the effective channel parameter
$q(D)=\left[1+\exp\!\left(8\pi(M-D)\omega\right)\right]^{-1}$
increases monotonically with the dilaton parameter $D$ at fixed $M$
and $\omega$. Corollary~\ref{cor:homogeneous-resource-ordering}
therefore predicts that the coherence in the homogeneous exterior
configuration is non-increasing with \(D\), consistent with the
corresponding coherence behavior reported in the cited study~\cite{liBosonicFermionicCoherence2024}. The homogeneous interior
configuration follows the opposite direction, whereas mixed
exterior--interior configurations are not constrained by
Theorem~\ref{thm:componentwise-multipartite-ordering}.

Related homogeneous-sector coherence trends have also been reported
for multipartite mixed-state inputs, states subjected to additional
fixed environmental noise, and other black-hole geometries
\cite{Liao2025Quantum,liuQuantumCoherenceQuantum2024,liEntropicUncertaintyCoherence2026}.
Together with the examples above, these studies provide further
instances where the homogeneous-sector coherence ordering is
consistent with previously reported relativistic coherence trends.

\subsection{Optimized Bell-functional values}

We next consider optimized Bell-functional values. Fix an \(m\)-party Bell scenario
\(\mathsf S\) associated with the partition
\(\mathcal P=B_1|\cdots|B_m\), together with fixed input and output alphabets
for each party \cite{brunnerBellNonlocality2014}.
Let $\mathcal B$ be a linear Bell functional defined in this scenario.
For a state $\rho$, its optimized Bell-functional value is
\begin{equation}
\begin{gathered}
  \beta_{\mathcal B}^{\mathsf S}(\rho)
  =
  \sup_{\bm M}
  \mathcal B[p_{\rho,\bm M}],
  \\
  p_{\rho,\bm M}(\bm a|\bm x)
  =
  \operatorname{Tr}
  \left[
    \rho
    \bigotimes_{j=1}^{m}
    M^{(j)}_{a_j|x_j}
  \right],
\end{gathered}
\end{equation}
where the supremum is taken over all local POVMs compatible with the
fixed input and output alphabets.
For either homogeneous sector, the corresponding intermediate map can
be regrouped according to the Bell parties as
$
  \mathcal V
  =
  \mathcal V_{B_1}\otimes\cdots\otimes\mathcal V_{B_m},
$
where each $\mathcal V_{B_j}$ is a deterministic local CPTP map acting
on the subsystems held by party $B_j$.
For any collection of local POVMs $\bm M$ performed on
$\mathcal V(\rho)$, define the pulled-back measurement operators by
$
  \widetilde M^{(j)}_{a_j|x_j}
  =
  \mathcal V_{B_j}^{\dagger}
  \left(
    M^{(j)}_{a_j|x_j}
  \right).
$
Since $\mathcal V_{B_j}$ is completely positive and trace preserving,
its adjoint is completely positive and unital. The operators
$\widetilde M^{(j)}_{a_j|x_j}$ therefore remain positive and satisfy
\begin{equation}
  \sum_{a_j}
  \widetilde M^{(j)}_{a_j|x_j}
  =
  \mathcal V_{B_j}^{\dagger}
  \left(
    \sum_{a_j}M^{(j)}_{a_j|x_j}
  \right)
  =
  \mathcal V_{B_j}^{\dagger}(I)
  =
  I,
\end{equation}
and hence form valid local POVMs on the input state.
The adjoint relation then gives
\begin{equation}
  p_{\mathcal V(\rho),\bm M}(\bm a|\bm x)
  =
  p_{\rho,\widetilde{\bm M}}(\bm a|\bm x).
\end{equation}
Consequently,
$
  \beta_{\mathcal B}^{\mathsf S}
  \bigl[\mathcal V(\rho)\bigr]
  \leq
  \beta_{\mathcal B}^{\mathsf S}(\rho),
$
so the optimized Bell-functional value is non-increasing under the
local product intermediate maps appearing in the homogeneous channel
ordering.

Therefore, for componentwise ordered parameter vectors
$\bm q\leq\bm q'$, Corollary~\ref{cor:homogeneous-resource-ordering}
gives
\begin{subequations}
\begin{equation}
\begin{aligned}
  \beta_{\mathcal B}^{\mathsf S}
  \bigl[\rho_{\mathrm{out}}(\bm q')\bigr]
  \leq
  \beta_{\mathcal B}^{\mathsf S}
  \bigl[\rho_{\mathrm{out}}(\bm q)\bigr],
\end{aligned}
\end{equation}
\begin{equation}
\begin{aligned}
  \beta_{\mathcal B}^{\mathsf S}
  \bigl[\rho_{\mathrm{in}}(\bm q')\bigr]
  \geq
  \beta_{\mathcal B}^{\mathsf S}
  \bigl[\rho_{\mathrm{in}}(\bm q)\bigr].
\end{aligned}
\end{equation}
\end{subequations}

Standard instances of
$\beta_{\mathcal B}^{\mathsf S}$ include the optimized Clauser--Horne--Shimony--Holt (CHSH),
Collins--Gisin--Linden--Massar--Popescu (CGLMP),
Mermin, and Svetlichny values, as well as the optimal winning
probability of a fixed nonlocal game
\cite{clauserProposedExperimentTest1969,collinsBellInequalitiesArbitrarily2002,merminExtremeQuantumEntanglement1990,svetlichnyDistinguishingThreebodyTwobody1987,cleveConsequencesLimitsNonlocal2004}.
For a fixed scenario $\mathsf S$ and Bell functional $\mathcal B$ with
local bound $\beta_{\mathcal B}^{\mathrm L}$, the violation excess
\mbox{\(\Delta_{\mathcal B}^{\mathsf S}(\rho)
=\max\{0,\beta_{\mathcal B}^{\mathsf S}(\rho)
-\beta_{\mathcal B}^{\mathrm L}\}\)}
is a non-decreasing function of
$\beta_{\mathcal B}^{\mathsf S}(\rho)$ and therefore inherits the same
exterior and interior ordering. This quantity characterizes the
violation of the chosen Bell functional in the fixed scenario, rather
than providing a faithful measure of all forms of Bell nonlocality
\cite{brunnerBellNonlocality2014}.

Bell-functional quantities have been investigated in Schwarzschild
spacetime for bipartite Dirac fields, open systems with additional
environmental noise, and tripartite initial states
\cite{xuProbingQuantumCorrelation2014,hePropertyVariousCorrelation2015,zhangHawkingEffectCan2023}.
For the homogeneous exterior and homogeneous interior configurations
considered in these works, specifying the dependence
$q=q(\lambda)$ on the relevant relativistic parameter $\lambda$
translates the channel ordering derived above into the corresponding
ordering of optimized Bell-functional values, in the same manner as
for the entanglement and coherence examples.
Since the parameter mapping proceeds analogously, the case-by-case
details are not repeated here. It should be mentioned that
mixed exterior--interior configurations are not covered by our present
theorem.

\subsection{Contractive correlation functionals from product marginals}
\label{sec:contractive-correlations}

Finally, we consider a class of correlation functionals constructed from
product-of-marginals reference states. Following the construction in Sec.~\ref{sec:entanglement-monotones},
we consider an arbitrary fixed partition
\(\mathcal P=B_1|\cdots|B_m\) and regroup the intermediate product map
accordingly as
\(\mathcal V=\bigotimes_{k=1}^{m}\mathcal V_{B_k}\).
Associated with this partition, define the product-of-marginals reference
state
$
  \Pi_{\mathcal P}(\rho)
  =
  \bigotimes_{k=1}^{m}\rho_{B_k}.
$
Because the channels acting on the complementary blocks are trace
preserving, the marginal of each block transforms as
$
  \bigl[\mathcal V(\rho)\bigr]_{B_k}
  =
  \mathcal V_{B_k}(\rho_{B_k}).
$
It follows that the product-of-marginals assignment is covariant under
\(\mathcal V\):
\begin{equation}
  \Pi_{\mathcal P}
  \bigl[\mathcal V(\rho)\bigr]
  =
  \mathcal V
  \bigl[\Pi_{\mathcal P}(\rho)\bigr].
  \label{eq:partition-reference-covariance}
\end{equation}
An explicit derivation of the above equation is given in the
\hyperref[app:product-reference]{Appendix}.

Let \(\mathfrak D\) be a state divergence satisfying the data-processing
inequality under CPTP maps. For the fixed partition
\(\mathcal P\), define the associated correlation functional
$
  T_{\mathfrak D}^{\mathcal P}(\rho)
  =
  \mathfrak D
  \left[
    \rho,
    \Pi_{\mathcal P}(\rho)
  \right].
$
Using the covariance relation
in Eq.~\eqref{eq:partition-reference-covariance} and the data-processing
inequality
\cite{nielsenQuantumComputationQuantum2010}, we obtain
\begin{align}
T_{\mathfrak D}^{\mathcal P}\bigl[\mathcal V(\rho)\bigr]
&=
\mathfrak D\bigl[
  \mathcal V(\rho),
  \Pi_{\mathcal P}\bigl(\mathcal V(\rho)\bigr)
\bigr]
\notag
\\
&=
\mathfrak D\bigl[
  \mathcal V(\rho),
  \mathcal V\bigl(\Pi_{\mathcal P}(\rho)\bigr)
\bigr]
\leq
T_{\mathfrak D}^{\mathcal P}(\rho).
\label{eq:correlation-contractivity}
\end{align}
Hence, \(T_{\mathfrak D}^{\mathcal P}\) is non-increasing under the
intermediate degrading maps and satisfies the monotonicity condition of
Corollary~\ref{cor:homogeneous-resource-ordering}. For parameter vectors
satisfying \(\bm q\leq\bm q'\) componentwise, the corollary therefore gives
\begin{subequations}
\begin{equation}
\begin{aligned}
    T_{\mathfrak D}^{\mathcal P}
\bigl[\rho_{\mathrm{out}}(\bm q')\bigr]
\leq
T_{\mathfrak D}^{\mathcal P}
\bigl[\rho_{\mathrm{out}}(\bm q)\bigr],
\end{aligned}
\end{equation}
\begin{equation}
\begin{aligned}
T_{\mathfrak D}^{\mathcal P}
\bigl[\rho_{\mathrm{in}}(\bm q')\bigr]
\geq
T_{\mathfrak D}^{\mathcal P}
\bigl[\rho_{\mathrm{in}}(\bm q)\bigr].
\end{aligned}
\end{equation}
\end{subequations}

A standard example is the Umegaki relative entropy
\(D(\rho\Vert\sigma)\)
\cite{umegakiConditionalExpectationOperator1962}, which satisfies the
data-processing inequality under CPTP maps
\cite{lindbladCompletelyPositiveMaps1975,uhlmannRelativeEntropyWignerYanaseDysonLieb1977}.
For a bipartition \(\mathcal P=X|\bar X\), the associated functional
\(T_D^{\mathcal P}\) is the mutual information \(I(X:\bar X)\), whereas
for the finest partition
\(\mathcal P=A_1|\cdots|A_N\), it is the multipartite total correlation
\(I(A_1:\cdots:A_N)\). More generally, it quantifies the total
correlation among the composite blocks specified by the fixed partition
\(\mathcal P\)
\cite{groismanQuantumClassicalTotal2005,liuCharacterizingCorrelationMultipartite2022}.
Representative fermionic relativistic quantum-information studies in
Schwarzschild, GHS, and multi-event-horizon spacetimes report
mutual-information trends consistent with the present homogeneous-sector
ordering
\cite{aliQuantumCharacteristicsEvent2024,tengBosonicFermionicMutual2026,liuQuantumPropertiesFermionic2023}.
The same construction also applies to other state divergences satisfying
the data-processing inequality, including the trace distance,
fidelity-based contractive distances such as the Bures distance, and
sandwiched R\'enyi divergences within their corresponding DPI parameter
ranges
\cite{gilchristDistanceMeasuresCompare2005,leditzkyDataProcessingSandwiched2017}.

Among these admissible constructions, we use the QJSD as the representative example in the
application below. The QJSD \cite{majteyJensenShannonDivergenceMeasure2005} is defined by
\begin{equation}
  \qjsd(\rho,\sigma)
  =
  S\!\left(\frac{\rho+\sigma}{2}\right)
  -\frac{1}{2}S(\rho)
  -\frac{1}{2}S(\sigma).
\end{equation}
Equivalently, writing
\(\mu=(\rho+\sigma)/2\), we have
$
  \qjsd(\rho,\sigma)
  =
  \frac{1}{2}D(\rho\Vert\mu)
  +
  \frac{1}{2}D(\sigma\Vert\mu).
$
The data-processing inequality for the relative entropy therefore implies
the CPTP contractivity of \(\qjsd\), and the monotonicity of the square
root implies the same property for
$
  \mathfrak D_{\mathrm{JS}}(\rho,\sigma)
  =
  \sqrt{\qjsd(\rho,\sigma)}.
$
For the finest partition, the resulting functional is the QJSD-based
collective coherence
\cite{radhakrishnanBasisindependentQuantumCoherence2019},
\begin{equation}
\resizebox{0.82\linewidth}{!}{$\displaystyle
  \Ccoll(\rho)
  =
  T_{\mathfrak D_{\mathrm{JS}}}^{A_1|\cdots|A_N}(\rho)
  =
  \sqrt{\qjsd(\rho,\pi_{\rho})},
  \quad
  \pi_{\rho}
  =
  \bigotimes_{i=1}^{N}\rho_{A_i}.
$}
\end{equation}
Consequently, \(\Ccoll\) inherits the homogeneous exterior and interior
horizon-channel orderings derived above. This is the quantity used in the
GHS application in the next section.

\section{Prediction of collective coherence in the GHS spacetime}
\label{sec:ghs}

We now apply the QJSD-based collective-coherence ordering derived above
to the GHS spacetime, a setting in which, to the best of our knowledge,
QJSD-based collective coherence has not previously been investigated. For a fermionic mode of frequency \(\omega\), the
effective horizon-channel parameter is
\(q(D,\omega,M)=\bigl[1+\mathrm e^{8\pi(M-D)\omega}\bigr]^{-1}\)\cite{wangQuantumDiscordMeasurementinduced2014a}.
In the physical region \(D<M\), \(q\) increases monotonically with \(D\)
at fixed \(M\) and \(\omega>0\).

As test inputs, we consider the noisy three-qubit GHZ and \(W\) families
$
  \rho_X(r)
  =
  r\proj{X}
  +
  \frac{1-r}{8}I_8,
$
where \(X\) denotes either the GHZ state
\(\ket{\mathrm{GHZ}}=(\ket{000}+\ket{111})/\sqrt{2}\)
or the \(W\) state
\(\ket{W}=(\ket{001}+\ket{010}+\ket{100})/\sqrt{3}\), with \(0\leq r\leq1\). These states provide inputs with different
multipartite correlation structures; neither the input family nor \(r\) affects the theorem-controlled
monotonic directions in the homogeneous configurations.
For the numerical illustration, we fix \(M=\omega=1\), with the plotted
endpoint \(D=1\) understood as the limiting case \(D\to M^{-}\). After subsystems
\(A\) and \(B\) undergo the horizon transformation while subsystem \(C\)
remains unchanged, the resulting output states for each input family
\(X\) are
\begin{subequations}
\label{eq:sector-states}
\begin{equation}
  \rho_X^{\mathrm{ee}}(D,r)
  =
  \left(
    \Eout{q}
    \otimes
    \Eout{q}
    \otimes
    \id_C
  \right)
  \!\left[\rho_X(r)\right],
  \label{eq:sector-state-ee}
\end{equation}
\begin{equation}
  \rho_X^{\mathrm{ii}}(D,r)
  =
  \left(
    \Ein{q}
    \otimes
    \Ein{q}
    \otimes
    \id_C
  \right)
  \!\left[\rho_X(r)\right],
  \label{eq:sector-state-ii}
\end{equation}
\begin{equation}
  \rho_X^{\mathrm{ei}}(D,r)
  =
  \left(
    \Eout{q}
    \otimes
    \Ein{q}
    \otimes
    \id_C
  \right)
  \!\left[\rho_X(r)\right].
  \label{eq:sector-state-ei}
\end{equation}
\end{subequations}
where \(\mathrm e\) and \(\mathrm i\) denote exterior and interior
outputs, respectively. The \(\mathrm{ee}\) and \(\mathrm{ii}\)
configurations are homogeneous, whereas \(\mathrm{ei}\) is mixed.
Because both input families are permutation symmetric, exchanging the
exterior and interior assignments of \(A\) and \(B\) gives no independent
case.

The result of Sec.~\ref{sec:contractive-correlations} directly gives, for
fixed \(r\) and \(D_2>D_1\),
\begin{subequations}
\begin{equation}
    \Ccoll\!\left[\rho_X^{\mathrm{ee}}(D_2,r)\right]
\leq
\Ccoll\!\left[\rho_X^{\mathrm{ee}}(D_1,r)\right],
\end{equation}
\begin{equation}
\Ccoll\!\left[\rho_X^{\mathrm{ii}}(D_2,r)\right]
\geq
\Ccoll\!\left[\rho_X^{\mathrm{ii}}(D_1,r)\right].
\end{equation}
\end{subequations}

No universal ordering follows for the mixed configuration
\(\rho_X^{\mathrm{ei}}\).

\begin{figure*}[t]
  \centering
  \includegraphics[width=\linewidth]{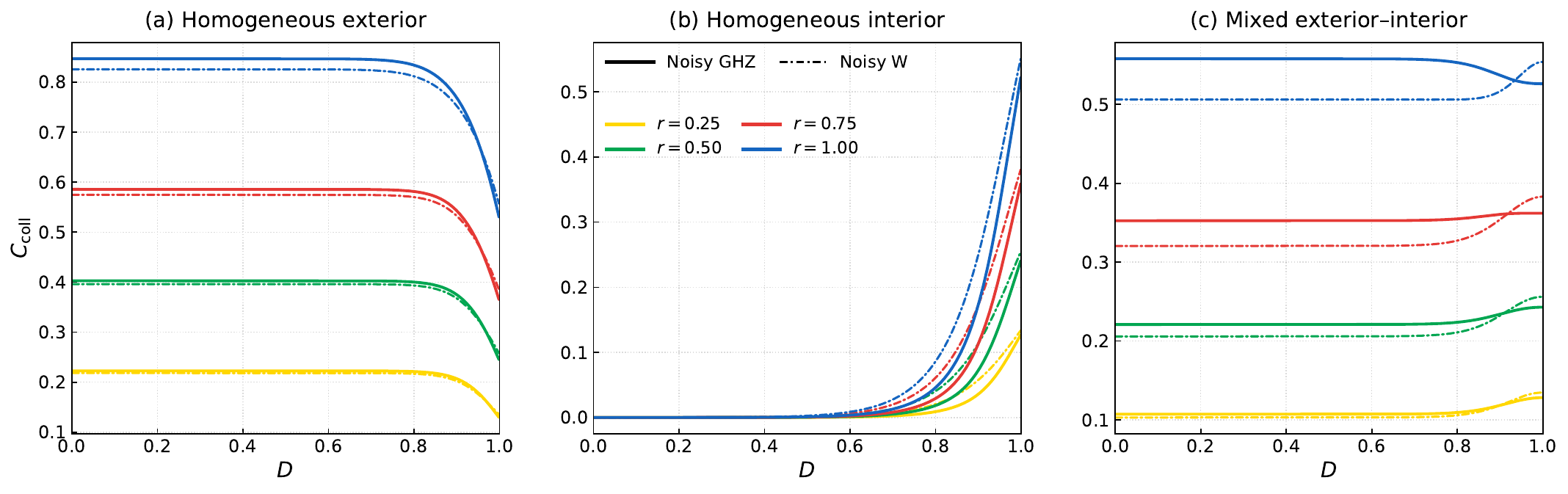}
  \caption{QJSD-based collective coherence for noisy GHZ and \(W\) inputs
in the GHS spacetime with \(M=\omega=1\). Subsystems \(A\) and \(B\)
undergo the horizon transformation, while subsystem \(C\) remains
unchanged. The left, middle, and right panels show the homogeneous
exterior (\(\mathrm{ee}\)), homogeneous interior (\(\mathrm{ii}\)), and
mixed exterior--interior (\(\mathrm{ei}\)) configurations, respectively. The endpoint \(D=1\) represents the limiting value \(D\to M^{-}\).}
  \label{fig:collective-coherence}
\end{figure*}

Figure~\ref{fig:collective-coherence} presents the corresponding numerical
results. For both input families and every displayed value of \(r\),
\(\Ccoll\) is non-increasing with \(D\) in the homogeneous exterior
configuration and non-decreasing in the homogeneous interior configuration,
in agreement with our theoretical predictions. By contrast, the mixed configuration exhibits input-dependent behavior.
Depending on the input family and the value of \(r\), the collective coherence
may increase, decrease, or display weak nonmonotonic variations with \(D\).
This contrast confirms that the mixed sector does not possess
the state-independent monotonic direction established
for the homogeneous sectors.

Together with the general results of Sec.~\ref{sec:resources}, this
application partially answers the question raised in the Introduction:
the recurring monotonic trends in homogeneous sectors have a common
channel-theoretic origin, whereas the behavior of mixed
exterior--interior sectors generally remains dependent on the input state.

\section{Discussion and conclusion}
\label{sec:discussion}

In summary, we have established exact and opposite post-processing orders for the
exterior and interior channels constructed from the effective
single-mode, two-level fermionic mode transformation. We find that as the relativistic
channel parameter \(q\) increases, the exterior channel becomes
progressively degraded, whereas the interior channel is ordered in the
reverse direction. Furthermore, for arbitrary multipartite input states, any choice of affected
subsystems, and nonuniform local parameters, these channel relations
imply opposite monotonic trends in homogeneous exterior and homogeneous
interior sectors for any state functional that is non-increasing under
the corresponding intermediate maps. Because the ordering is established at the channel level, the same
argument applies uniformly across different input states, admissible
resource and correlation functionals. Moreover, the
application to QJSD-based collective coherence in the GHS spacetime
illustrates both the predicted monotonicity in the homogeneous sectors
and the input-dependent behavior of mixed exterior--interior
configurations. 

As for the scope and limitations of our work, it should be noted that at the field-theoretic level, the present construction adopts an
effective local, single-mode, two-level occupation-number description.
The single-mode approximation is not valid for arbitrary field states,
although it can be justified for suitably localized wave packets under
appropriate peaking conditions
\cite{bruschiUnruhEffectQuantum2010}.
A multimode treatment may therefore modify the mode decomposition and
the resulting product-channel structure. The internal spin degree of
freedom is also not explicitly resolved here. Spin-resolved Dirac modes
require larger local Fock spaces containing vacuum, spin-up, spin-down,
and double-occupation sectors, and consequently lead to
higher-dimensional complementary channels
\cite{leonSpinOccupationNumber2009}.
Likewise, bosonic fields involve an infinite-dimensional occupation
structure and require a separate analysis of their channel-ordering
properties
\cite{panHawkingRadiationEntanglement2008,monteroEntanglementArbitrarySpin2011}.
Finally, the effective mode-qubit representation used throughout this
work does not constitute a fully algebraic treatment of fermionic
subsystems. Such a treatment would additionally account for fermionic
mode embeddings, operator ordering, partial traces, and parity
superselection
\cite{friisFermionicmodeEntanglementQuantum2013,szalayFermionicSystemsQuantum2021}. On the other hand, at the resource-theoretic level, the exact post-processing identities do
not order arbitrary state quantities. They imply monotonicity only when
the functional under consideration is non-increasing under the relevant
intermediate CPTP map. This is precisely the condition used in
Sec.~\ref{sec:resources}; quantities that do not satisfy this condition
are not covered by the present argument.

These limitations also suggest several directions for extending the
present framework, including spin-resolved and multimode horizon
transformations. A related channel-based treatment has recently been
developed for bosonic noninertial dynamics within a general
quantum-resource-theoretic framework~\cite{harikrishnanInfluenceNoninertialDynamics2026}. Whether the
channel-ordering structure established here persists for bosonic horizon
channels remains to be examined. Another direction is to characterize
the conditions under which fermionic horizon channels preserve
particular quantum resources.

Overall, within the homogeneous sectors, the framework explains why
recurring monotonic trends arise from a common channel-ordering
structure: the physical setting determines the parameterization of
\(q\), whereas the resource-theoretic monotonicity condition determines
which output-state quantities inherit that order.
More broadly, this channel-ordering viewpoint points toward a shift in relativistic quantum information away from case-by-case calculations to a structural theory of how spacetime-induced dynamics constrain the flow and redistribution of quantum resources. Its extension to multimode, spin-resolved, and bosonic settings may ultimately provide a unified framework for classifying relativistic resource dynamics and for identifying operational quantum-information signatures of horizons and curved spacetime.

\section*{Acknowledgments} This work is supported by the Natural Science Foundation of Shanghai, China under Grant No. 25ZR1401098, the National Natural Science Foundation of China under Grants No. 92265209, and the Shanghai Municipal Science and Technology Major Project under Grant No. 2019SHZDZX01.

\begin{appendices}

\refstepcounter{section}
\setcounter{equation}{0}
\renewcommand{\theequation}{A\arabic{equation}}

\section*{Appendix}
\label{app:product-reference}

We fix an arbitrary partition
\(\mathcal P=B_1|\cdots|B_m\), and regroup the local product CPTP map as
\(\mathcal V=\bigotimes_{k=1}^{m}\mathcal V_{B_k}\), where
\(\mathcal V_{B_k}\) contains all local channels acting on the
subsystems in \(B_k\). Channels acting on unaffected subsystems are
understood to be identity channels.

For each block \(B_k\), let \(\overline{B_k}\) denote its complement and
write
\(\mathcal V_{\overline{B_k}}
=\bigotimes_{\ell\neq k}\mathcal V_{B_\ell}\).
For an arbitrary operator \(X_{B_k}\) on \(B_k\), we have
\begin{align}
\operatorname{Tr}\!\left[
  X_{B_k}\bigl[\mathcal V(\rho)\bigr]_{B_k}
\right]
&=
\operatorname{Tr}\!\left[
  \left(X_{B_k}\otimes I_{\overline{B_k}}\right)
  \left(
    \mathcal V_{B_k}\otimes
    \mathcal V_{\overline{B_k}}
  \right)(\rho)
\right]
\notag
\\
&=
\operatorname{Tr}\!\left[
  \left(
    \mathcal V_{B_k}^{\dagger}(X_{B_k})
    \otimes
    \mathcal V_{\overline{B_k}}^{\dagger}
    (I_{\overline{B_k}})
  \right)\rho
\right]
\notag
\\
&=
\operatorname{Tr}\!\left[
  \mathcal V_{B_k}^{\dagger}(X_{B_k})
  \rho_{B_k}
\right]
\notag
\\
&=
\operatorname{Tr}\!\left[
  X_{B_k}\mathcal V_{B_k}(\rho_{B_k})
\right].
\label{eq:appendix-block-reduction}
\end{align}
Here we used the trace preservation of
\(\mathcal V_{\overline{B_k}}\), which implies
\(\mathcal V_{\overline{B_k}}^{\dagger}
(I_{\overline{B_k}})=I_{\overline{B_k}}\).
Since this equality holds for every \(X_{B_k}\), it follows that
$
\bigl[\mathcal V(\rho)\bigr]_{B_k}
=
\mathcal V_{B_k}(\rho_{B_k}).
$

For the product-of-marginals reference state
\(\Pi_{\mathcal P}(\rho)=\bigotimes_{k=1}^{m}\rho_{B_k}\), we therefore
obtain
\begin{align}
\Pi_{\mathcal P}\bigl[\mathcal V(\rho)\bigr]
&=
\bigotimes_{k=1}^{m}
\bigl[\mathcal V(\rho)\bigr]_{B_k}
=
\bigotimes_{k=1}^{m}
\mathcal V_{B_k}(\rho_{B_k})
\notag
\\
&=
\mathcal V\!\left(
  \bigotimes_{k=1}^{m}\rho_{B_k}
\right)
=
\mathcal V\bigl[\Pi_{\mathcal P}(\rho)\bigr].
\end{align}
Hence the product-reference assignment is covariant under every local
product CPTP map.

\end{appendices}

\backmatter

\bibliography{references_epjc}
\end{document}